# COVID-19 Imposes Rethinking of Conferencing – Environmental Impact Assessment of Artificial Intelligence Conferences

Pavlina Mitsou, Nikoleta-Victoria Tsakalidou, Eleni Vrochidou[0000-0002-0148-8592], and George A. Papakostas[0000-0001-5545-1499]

MLV Research Group, Department of Computer Science, International Hellenic University, 65404 Kavala, Greece
gpapak@cs.ihu.gr

**Abstract.** It has been noticed that through COVID-19 greenhouse gas emissions had a sudden reduction. Based on this significant observation, we decided to conduct a research to quantify the impact of scientific conferences' air-travelling, explore and suggest alternative ways for greener conferences to reduce the global carbon footprint. Specifically, we focused on the most popular conferences for the Artificial Intelligence community based on their scientific impact factor, their scale, and the well-organized proceedings towards measuring the impact of air travelling participation. This is the first time that systematic quantification of a state-of-the-art subject like Artificial Intelligence takes place to define its conferencing footprint in the broader frames of environmental awareness. Our findings highlight that the virtual way is the first on the list of green conferences' conduction although there are serious concerns about it. Alternatives to optimal conferences' location selection have demonstrated savings on air-travelling CO2 emissions of up to 63.9%.

**Keywords:** Artificial intelligence, COVID-19, environmental sustainability, carbon footprint, carbon emission assessment, green economy, green AI.

## 1 Introduction

COVID-19 has affected all aspects of our daily lives such as the way of working, education, entertainment, travel and mobility patterns, also the way that scientific conferences are conducted. People try to find efficient alternatives to interact and collaborate with each other in the "new normal way". According to Jean-Eric Paquet, the European Commission's Director General for Research and Innovation, what is happening in the research community is that new opportunities are emerging [1]; by discovering new ways of working, closer to the digital world, new incentives are created for scientists to work in different places, outside of their lab or field of interest, since in this convenient remote way, everyone is more available. Therefore, participants' diversity is facilitated by online conferencing, and novel holistic and inter- and transdisciplinary approaches in research are emerging [2].

Prior to the pandemic, researchers would air-travel to participate in conferences and scientific research projects, contributing to a significant amount of $CO_2$ emissions; environmental ethical issues, therefore, rise, regarding the environmental impact of international conferencing [3]. As awful as this pandemic is, climate change could be worse [4]. It has been noted that through COVID-19 the greenhouse gas emissions as well as air pollutants had a sudden reduction [5]; in 2020, a reduction of 8% in emissions was estimated by the International Energy Agency, translated in 4 billion tonnes of less carbon ($CO_2$) [6]. Transport accounts for about 24% of global $CO_2$ emissions [7]. More specifically, UC Santa Barbara (UCSB) calculated its $CO_2$ footprint and concluded that almost one third of it came from staff and teachers traveling to meetings, talks and conferences [8], equal to a Philippines city of 27,500 people annually carbon footprint.

Many people perceived the COVID-19 positive impact in the natural environment and changed their daily habits regarding energy savings, waste management, mobility, and other environment-related behaviors [9]. Therefore, scientific conferences pose a dilemma for climate-conscious researchers [10]. Is it environmentally ethical for research to pay such an environmental price to participate in international conferences? Not participating at all could possibly halt research or harm it in any way?

Towards this end, eco-friendly conference models to reduce $CO_2$ emissions are suggested. One way is to optimize meeting locations in order to minimize greenhouse gas emissions [11, 12]. By choosing the meeting place close to the participants, therefore by minimizing travel distances, estimated a reduction in carbon emissions by 0.2 tonnes per person [11]. United Nations Framework Convention on Climate Change (UNFCCC) proposed specific actions to reduce the carbon footprint of conferences [13], regarding alternatives in traveling, accommodation, etc. Erasmus+ also promoted sustainable means of transportation and environmentally responsible behaviors [14]. However, the best prevention is abstinence; researchers should challenge themselves if their presence is mandatory. Cutting back on conference travelling and going virtual could help, as COVID-19 traveling restrictions already have proven.

However, someone could argue that online conferences do not have the same impact compared to the on-site; Internet is not able to re-create a realistic environment for researchers to interact, develop ideas and collaborations. But is this entirely true?

Online conferencing can remove barriers and facilitate the participation of people unable to attend in-person meetings, due to physical, social, religious, ethical, family, economic or other issues [15]. Prior to the COVID-19 era, certain international academic meetings had already started experimenting with alternative participation ways, with the November meeting of the European Biological Rhythms Society (EBRS) being the first to adopt a systematic approach of offering virtual hubs across 18 time zones. The organizers called psychologists to assess whether organizational techniques and technology could help networking and interaction. Ludwig Maximilian University (LMU) psychologist Anne Frenzel assessed that there were several advantages to virtual meetings, such as the reduction of emissions, the gain of time and energy for travelers as well as the ability for students to freely attend. The most important advantage, however, was the release from the bureaucracy to which the researchers were subjected to be able to move from their institute and air-travel to attend conferences abroad [16]. Moreover, recorded on-line presentations enable asynchronous attendance and therefore greater and broader accessibility to a larger and more diverse audience [17]. Last but not least, several researchers mentioned that through virtual conferences they approached a number of people in the online chat with questions [15, 18]; through the chat, it seemed easier to approach people by fighting the fear of socializing and the embarrassment about the use of English and correct pronunciation [17]. Statistics speak for themselves; in 2019 the European Geosciences Union (EGU) Assembly in Vienna had 16,200 registrations, while the digital General Assembly in 2020 had 10,000 more individual participants [18]. The online portal has enabled many researchers to attend more conferences than it would be possible to participate in physical presence. Nature conducted a poll among its readers where 74% stated that scientific meetings should continue to go virtual even after COVID-19 [15]. Moreover, 75% answered that they had attended several online conferences since March 2021, when conferences supported hybrid mode with physical on-site and on-line presentations, and 18% had attended at least one.

On the other hand, sitting for hours behind a computer screen and listening to live or pre-recorded lectures without being able to step out of the daily routine, does not sound appealing. Networking was also affected; it seems that online events put barriers to the connection of some people in the network who could possibly serve as future collaborators [15]. One more drawback is the different time zones, making it difficult for organizers and participants to go live due to time-zone related conflicts, i.e., night-time, business hours, etc. Technological issues such as connection unreliability, poor internet and power outages were the major drawbacks of all; an associate professor at the University of the Andes said that almost had to cancel his talks at Botany 2020, due to such issues [17]. Despite the reported drawbacks, virtual conferences overall seem to be a good experience and that they will trigger a huge change in conferences' conduction in the future [18, 19]. One of the biggest beneficiaries is the environment; virtual conference organizers in the USA estimated that total $CO_2$ emissions were less than 1% of a traditional «fly-in» event [18, 20].

After considering all the above in conjunction with the emerging changes that COVID-19 brought up to the surface in the last two years, we considered measuring the carbon footprint of Artificial Intelligence conferences, from 2011 to 2021, covering eight pre-COVID years and two during the pandemic. Conferencing air-travelling $CO_2$ emissions have been calculated for six selected Artificial Intelligence conferences, along with the emission savings by optimally selecting the conference location in two different proposed ways. This work is the first systematic quantification of the emissions related to a state-of-the-art subject like Artificial Intelligence to define its conferencing footprint towards environmental awareness.

The environmental impact of academic conferences has been a subject of interest in the post-COVID-19 era. The carbon footprint of presenters' air traveling to pre-COVID pediatric urology conferences was investigated, considering seven related conferences [21]. Results indicated travel-related carbon emissions and suggested exploring novel conference strategies. The reduction of academic flying due to COVID-19 was investigated in [22]; emissions, however, were not quantified, and the research that was based on questionnaires, determined the basic factors that were essential for academics to substitute physical presence with digital. Scholars' experiences with digital conferencing during the pandemic was also investigated by other researchers [23, 24]. A quantification of the transition from in-person conferencing to virtual was provided by [25]; yet, no specific conferences were examined. The research focused on quantifying and comparing the life cycle environmental impact of in-person, virtual and hybrid conferencing, including the carbon footprint of transportation, catering, conference execution and accommodation. Travel emissions of three global conferences of the International Society for Industrial Ecology were analyzed in [26], towards modeling reduction potentials. The carbon footprint of a multination-al knowledge organization in the service sector was measured in [27] and alternative emission scenarios for post-COVID world were examined.

Compared to the above related works of the recent literature, the contribution and innovations of this work can be summarized in the following distinct points:

1. This work presents for the first time in the international literature, both quantitatively and qualitatively, the environmental footprint of international conferences in the field of artificial intelligence.
2. This specific work is based on an original research methodology implemented in software based on specific considerations/assumptions in order to be able to quantify the environmental footprint.
3. The work concludes with some proposals aiming to problematize the scientific community and raise awareness among conference organizers to find more environmentally optimal ways of conducting conferences.
4. An important feature of the proposed methodology is its high degree of adaptation since it is not limited to the field of artificial intelligence but can be extended and applied to any scientific discipline.

In addition, in the background of the aforementioned basic research, this work aims (1) to highlight the need for the online conferencing alternative based on $CO_2$ air-travelling emissions quantification, (2) to record which countries are scientifically active in Artificial Intelligence based on conferences participations and (3) to investigate the emissions that certain Artificial Intelligence conferences generate so as to propose greener alternatives by selecting the conference location based on the activity and on the participants' origins. The outcomes of this work could provide useful insights, impose rethinking of conferencing, propose greener alternatives and raise concerns to the research community on the way of networking in the future.

The rest of the paper is structured as follows. Materials and methods used in this work are presented in Section 2. Experimental results are summarized in Section 3. In Section 4, research findings are discussed in detail. Section 5 concludes the paper.

## 2 Materials and Methods

### 2.1 Selected Conferences

To quantify the impact of the air pollution due to the flights of conference travelling, certain assumptions and decisions were made so as to clarify the properties of our problem. Our study was based on the conferences between 2011-2021, which met the three following criteria: (1) their impact factor was high based on the h-index ranking, (2) the scale of the conference was large and (3) the information of proceedings was organized in a clear way to automatically extract all necessary information. The main source of data extraction for the publications was the Elsevier Scopus API [28] since it is the largest abstract and citation database of peer-reviewed literature. By using the conference name and year, Scopus returned the number of publications and all the information for each publication. Information that could not be retrieved from Scopus was not included.

The selected case study of Artificial Intelligence conferences is due to their interdisciplinary interest, including both machine learning and artificial intelligence, being currently at the forefront of scientific innovations. These conferences were the following: (1) European Conference on Computer Vision (ECCV), held every two years, (2)

International Conference on Machine Learning (ICML), (3) International Joint Conference on Neural Networks (IJCNN), (4) IEEE Congress on Evolutionary Computation (IEEE CEC), (5) IEEE CVF Conference on Computer Vision and Pattern Recognition (IEEE CVPR), (6) IEEE International Conference on Robotics and Automation (IEEE ICRA). Information regarding the selected conferences is included in Table 1. It should be noted that some of the most important Artificial Intelligence conferences were not included since they did not have well-organized proceedings in a way as to extract the necessary information automatically, such as the Conference on Neural Information Processing Systems (NIPS), the International Conference on Learning Representations (ICLR), the International Joint Conferences on Artificial Intelligence (IJCAI), the Association for the Advancement of Artificial Intelligence (AAAI) and more. Due to the same reason, workshops were not considered even though they form a major part of big AI conferences, permitting locals to attend so might "dilute" the per person emissions. Additionally, due to not well-structured proceedings, the data on IEEE CVPR before 2018 could not be retrieved. Missing data may leave a concern of unfair comparison; however, the focus is to reveal a trend, therefore absolute values are of no particular purpose. Moreover, IEEE CVPR is a very large conference and the data from 2018 to 2021 are so many that are not considered to limit the research.

**Table 1.** Information regarding the selected conferences held between 2011-2021.

| Conf. Name | Included Years | Comments |
|---|---|---|
| ECCV | 2012, 2014, 2016, 2018, 2020 | Held every two years |
| ICML | 2011-2021 | Held every year |
| IJCNN | 2011-2021 | Held every year |
| IEEE CEC | 2011-2021 | Held every year |
| IEEE CVPR | 2018-2021 | Held every year, years before 2018 could not be automatically retrieved due to not well-structured proceedings |
| IEEE ICRA | 2011-2021 | Held every year |

## 2.2 Considerations and Research Assumptions

In order to quantify our research, we made some assumptions. First, we assumed that from all the authors of each publication, only one author would travel to the country where the conference was held and that she/he would start travelling from the affiliation city of the paper. Second, we assumed that the first author would air-travel in the optimal way and would opt for the economy class. By optimal, it means that she/he would travel from the nearest airport from her/his affiliation location to the nearest conference's airport. In order to find the nearest airport for both the affiliation town and the conference location, we used the travelpayouts [29], which returned the International Air Transport Association airport (IATA) code of the nearest airport requested by the city name.

For the $CO_2$ calculations, the GoClimate API [30] is used, which calculated the carbon footprint round trip of each participant of the conferences. GoClimate calculates the footprints based on the great circle distance and by using the Haversine formula. For non-direct flights, route segments were used as individual flights.

It should be noted that in our calculations we do not consider any people that traveled and attended the conference, other than the first authors of all presented papers. Therefore, the results of the calculations on the $CO_2$ emissions would be higher if those participants could have also been considered.

### 2.3 Optimal Conference Location Selection Approaches

In our experiments, we estimated the carbon emissions for conference travelling from 2011 - 2019 and the savings from 2020 and 2021 due to COVID-19 travel restrictions.

In order to estimate the potential savings, we investigated the cases of a conference location determination based on two specific criteria:

The first criterion was related to the optimal location based on all countries. Therefore, we assumed that every single country world-wide was eligible to be the conference organizer and we tried to find the optimal location by summing up all the absolute distances from each country to all of the paper affiliation locations. The country with the least sum was elected to be the conference organizer. From the elected country, we assumed that the capital would organize the conference and we calculated the actual distance from the biggest international airport of the capital to each of the paper's affiliation location, so as to calculate the respective $CO_2$ emissions. The flow of the «optimal location based on all countries» algorithm is illustrated in Fig. 1(a). For example, assume only two paper submissions to a conference; the sum of distances of the two papers' affiliation locations and each world capital separately, is calculated. The minimum sum of distances indicates the capital that is cumulatively closer to both paper submissions and therefore it is selected as the optimal conference organizer based on this approach.

The second criterion was related to the optimal location based on the country with the most paper submissions. In this way, the majority of the conference participants would not have to air-travel. After the determination of the country with the most paper submissions, the same steps were followed as before in order to calculate the $CO_2$ emissions. The flow of the «optimal location based on paper submissions» algorithm is illustrated in Fig. 1(b). For example, assume three paper submissions to a conference: two from China and one from Germany. The maximum sum of papers with the same affiliation country indicates the country that cumulatively holds the majority of submissions and therefore its capital is selected as the optimal conference organizer based in this approach.

### 2.4 Overall Proposed Methodology

Fig. 2 illustrates the flow of the overall proposed methodology; first, we used the Elsevier's Scopus API to extract the first author's location from all conference papers. Then we used the Travelpayours API to define the nearest airport for each extracted location of the previous step. At the third step of the proposed methodology, we employed GoClimate API to calculate the carbon footprint round trip of each participant (1) for the actual conference location, (2) for the optimal location based on the minimum distance between all countries and (3) for the optimal location based on paper submissions. The reported results are the cumulative results of the above process per conference.

## 3    Results

For our case study, we focused on the most popular conferences on Artificial Intelligence based on their scientific impact factor, scale and the well-organized proceedings, found in Scopus. These conferences were the ECCV, the ICML, the IJCNN, the IEEE CEC, the IEEE CVPR and the IEEE ICRA.

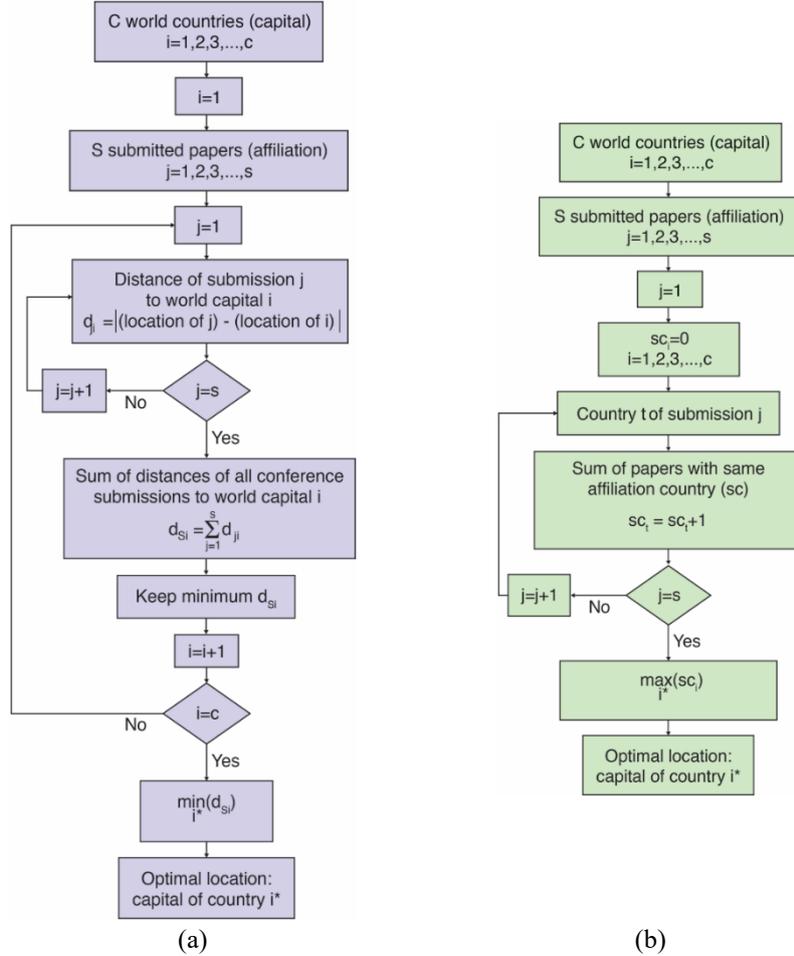

(a)                                          (b)

**Fig. 1.** Flow of: (a) the «optimal location based on all countries» algorithm, (b) the «optimal location based on paper submissions» algorithm.

The data from the above conferences were used to estimate the savings on carbon footprint, if participants had chosen to attend virtually instead of flying to the conference location. Moreover, we calculated the savings on carbon emission if the optimal conference locations had been selected, in two different ways. First, we assumed that all countries were eligible to organize the conference and second, the organizer was the country with the most reported submissions. The results for before

COVID-19 years, from 2011 to 2019, are included in Table 2, while for the COVID-19 years, from 2020 to 2021, are included in Table 3. It should be noted that in 2021, most of the conferences went virtual and therefore no conference locations are considered. Negative savings on the Tables refers to extra $CO_2$ gas emissions. Optimal location does not always mean lower emissions, since optimal location determination is based on assumptions. For example, even if most of the submissions belong to the organizing country, air-travelling emissions may be high, if the rest of the submitted papers come from distant countries. For this reason, negative savings are only noticed when the conference location selection is based on paper submissions. When the conference location is based on the minimum distance between all countries, then the selection of the organizing city is assumed to be the capital of the county. In this case, if the airport near the actual conference location is not the same with the capital's airport, emissions can be reported, either positive or negative. Savings of gas emissions of up to 63.9% are reported, when the optimal conference location is selected based on the affiliation of most paper submissions. Results indicate that eco-friendly conference models are indeed possible.

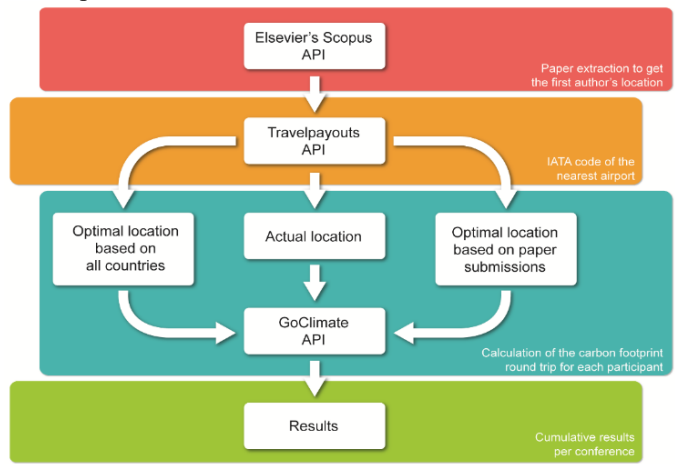

**Fig. 2.** Flow of the proposed methodology.

During the COVID-19 pandemic, most of the conferences went virtual. In the case of fully virtual conferences, savings on $CO_2$ emissions were 100%, since no one travelled to attend. For hybrid conferences, there was no input on which papers were presented on-line and in-person so as to calculate the savings on air-travelling emissions. However, in all cases, a location was announced prior to the decision for the conference to go fully virtual/hybrid. Therefore, in order to investigate these cases as well, the calculations were done in the same way as before, assuming that the conferences were all attended in-person at the pre-announced locations. Results indicated that, in these cases, $CO_2$ savings could have been achieved, fluctuating between 100%, if the conference was fully virtual, and a calculated value of $CO_2$ savings (indicated in Table 3), if the conference was conducted with a physical appearance for all participants. Hybrid way $CO_2$ savings would range between these two extreme values. Fig. 3 illustrates the CO2 emissions in tonnes in the past 10 years (2011-2021) due to conference flying for attending the six selected Artificial Intelligence conferences at their actual locations. Although 2020 and 2021 conferences were mostly virtual, the calculations include their emissions if they had been in-person. From Fig. 3 is obvious that the total emissions are higher in the most recent years. The latter is due to the cumulative contribution of IEEE CVPR, not included in the years prior 2018. Higher emissions are noted for ICML, especially when located in the US, such as the ICML 2013 and ICML 2019.

Fig. 4 includes the same information, if the conferences were conducted in the optimal location based on all countries (Fig. 4a) or on paper submissions (Fig. 4b). The latter is to better visualize the reduction of air-travelling CO2 emissions due to optimal locations' selection. As is it can be seen in Fig. 4a, if the location was selected based on minimum distance from all participants' countries, all emissions in all years would have been reduced, compared to the actual locations depicted in Figure 1. Notable is the difference in year 2013, when from 18K tonnes, the emissions due to the proposed approach could have been reduced to around 7K tonnes. Fig. 4b depicts the gas emissions if the conference location was selected based on the location/origin of most of the authors who submitted their papers.

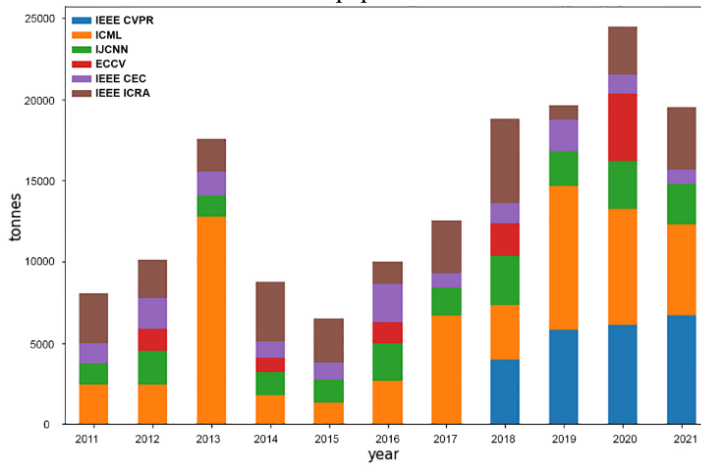

**Fig. 3.** Total CO2 emission tonnes per conference per year (actual location).

This alternative also results to reduced emissions, compared to the actual location depicted in Fi. 1. At this point, it should be noted that conference participation during years was not affected by the COVID-19 pandemic. On the contrary, more submissions were reported in most of the conferences in 2020 and 2021, as it can be seen in Fig. 5. The latter can be attributed mainly to financial reasons (savings on registration fees, accommodation and transportation).

In what follows, the results of applying the proposed methodology are presented. The results before COVID-19 years (2011 to 2019) are summarized in Table II, while the results during COVID-19 years (2020 to 2021) are summarized in Table III. More specifically, in both tables the following details are included:

   • The actual location where the conference took place.

   • The optimal location of the conference based on the minimum distance between all countries (BOC) that have submitted papers.

   • The corresponding savings in $CO_2$ emissions for air-travelling if the conference was conducted in the optimal location BOC.

   • The optimal location of the conference based on the location/origin of most of the authors who submitted their papers (BPS).

   • The corresponding savings in $CO_2$ emissions for air-travelling if the conference was conducted in the optimal location BOC.

All conclusions presented and discussed in the next section arise from the observation of the results contained in these two tables. Comparative study of the savings in $CO_2$ based on the two proposed alternatives (BOC and BPS) can lead to valuable conclusions about which approach could be more efficient and ecological.

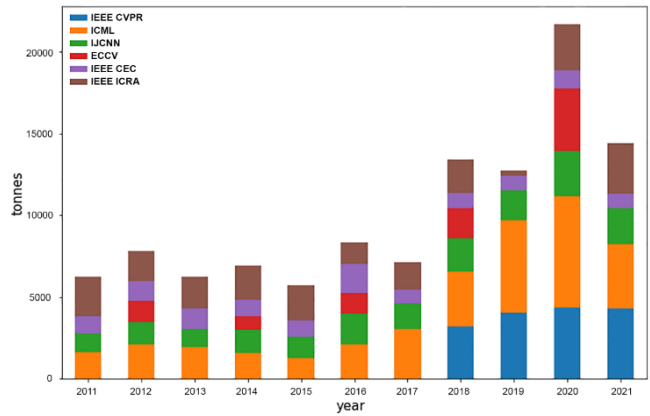

(a)

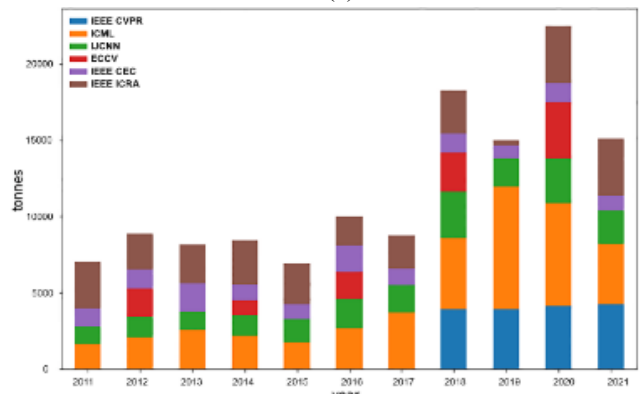

(b)

**Fig. 4.** Total $CO_2$ emission tonnes per conference per year if the conference was conducted in the optimal location based on: (a) all countries, (b) on paper submissions.

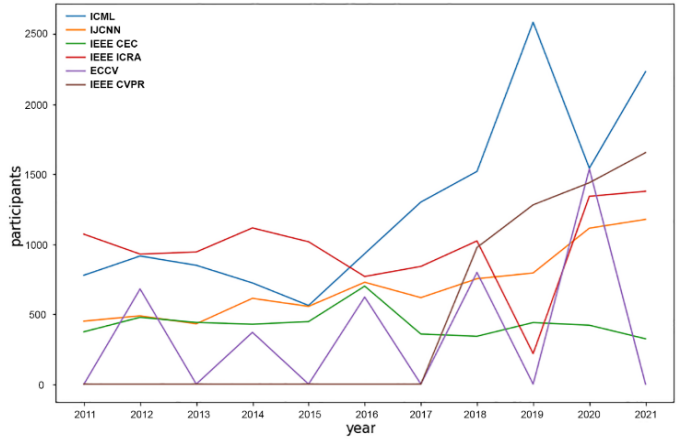

**Fig. 5.** Total CO2 emission tonnes per conference per year if the conference was conducted in the optimal location based on: (a) all countries, (b) on paper submissions.

Aggregating actual conference locations in the tables, can give us insight into trends in conference location selection. Moreover, the selection of the place based on the origin of the authors from the respective article submissions, can give us information about how the research interest in AI is distributed worldwide. However, the focus is on the savings in CO2 based on the two proposed approaches.

**Table 2.** Actual conference locations (country, town, airport) and optimal locations based on all countries (BOC) and based on paper submissions (BPS). Savings (%) in CO2 emissions for conference air-travelling to optimal locations, 2011 to 2019 (before COVID-19).

| Conf. Name | Actual Location | Optimal location BOC | Savings in CO2 - location BOC | Optimal location BPS | Savings in CO2 - location BPS |
|---|---|---|---|---|---|
| **2011** | | | | | |
| ICML | United States (Bellevue, Washington, LKE) | China (Beijing, PEK) | 33.0% | China (Beijing, PEK) | 33.0% |
| IJCNN | United States (San Jose, California, SJC) | United States (Washington, JFK) | 10.2% | United States (Washington, JFK) | 10.2% |
| IEEE CEC | United States (New Orleans, Louisiana, MSY) | Norway (Oslo, OSL) | 18.8% | United States (Washington, JFK) | 6.6% |
| IEEE ICRA | China (Shanghai, PVG) | Norway (Oslo, OSL) | 21.9% | United States (Washington, JFK) | 0.6% |
| **2012** | | | | | |
| ECCV | Italy (Florence, FLR) | United Kingdom (London, LHR) | 3.8% | United States (Washington, JFK) | -32.4% |
| ICML | United Kingdom (Edinburgh, Scotland, EDI) | China (Beijing, PEK) | 14.3% | China (Beijing, PEK) | 14.3% |
| IJCNN | Australia (Brisbane, BNE) | Mongolia (Ulaanbaatar, UBN) | 34.0% | China (Beijing, PEK) | 33.6% |
| IEEE CEC | Australia (Brisbane, BNE) | China (Beijing, PEK) | 33.8% | China (Beijing, PEK) | 33.8% |
| IEEE ICRA | United States (St Paul, Minnesota, STP) | Norway (Oslo, OSL) | 23.5% | United States (Washington, JFK) | 1.8% |
| **2013** | | | | | |

| | | | | |
|---|---|---|---|---|
| ICML | United States (Atlanta, ATL) | Finland (Helsinki, HEL) | 28.9% | United States (Washington, JFK) | 6.6% |
| IJCNN | United States (Dallas, Texas, DFW) | United Kingdom (London, LHR) | 17.6% | United States (Washington, JFK) | 7.2% |
| IEEE CEC | Mexico (Cancun, CUN) | United Kingdom (London, LHR) | 17.1% | Brazil (Brasilia, BSB) | -16.1 % |
| IEEE ICRA | Germany (Karlsruhe, FKB) | Norway (Oslo, OSL) | 1.1% | United States (Washington, JFK) | -28.8% |
| **2014** | | | | | |
| ECCV | Switzerland (Zurich, ZRH) | Iceland (Reykjavik, KEF) | 3.3% | United States (Washington, JFK) | -9.5% |
| ICML | China (Beijing, PEK) | Russia (Moscow, SVO) | 11.2% | United States (Washington, JFK) | -21% |
| IJCNN | China (Beijing, PEK) | China (Beijing, PEK) | 0.0% | China (Beijing, PEK) | 0.0% |
| IEEE CEC | China (Beijing, PEK) | China (Beijing, PEK) | 0.0% | China (Beijing, PEK) | 0.0% |
| IEEE ICRA | China (Hong Kong, HKG) | Denmark (Copenhagen, CPH) | 43.1% | United States (Washington, JFK) | 21.2% |
| **2015** | | | | | |
| ICML | France (Lille, LIL) | Finland (Helsinki, HEL) | 5.3% | United States (Washington, JFK) | -31.5% |
| IJCNN | Ireland (Killarney, KIR) | Denmark (Copenhagen, CPH) | 6.2% | China (Beijing, PEK) | -9.8% |
| IEEE CEC | Japan (Sendai, SDJ) | China (Beijing, PEK) | 9.1% | China (Beijing, PEK) | 9.1% |
| IEEE ICRA | United States (Seattle, Washington, DCA) | Iceland (Reykjavik, KEF) | 19.6% | United States (Washington, JFK) | 2.0% |
| **2016** | | | | | |
| ECCV | Netherlands (Amsterdam, AMS) | Norway (Oslo, OSL) | 0.6% | United States (Washington, JFK) | -38.5% |
| ICML | United States (New York, JFK) | Norway (Oslo, OSL) | 21.4% | United States (Washington, JFK) | 0.0% |
| IJCNN | Canada (Vancouver, YVR) | Mongolia (Ulaanbaatar, UBN) | 18.0% | China (Beijing, PEK) | 17.2% |
| IEEE CEC | Canada (Vancouver, YVR) | China (Beijing, PEK) | 24.9% | China (Beijing, PEK) | 24.9% |
| IEEE ICRA | Sweden (Stockholm, ARN) | Netherlands (Amsterdam, AMS) | 3.2% | United States (Washington, JFK) | -35.6% |
| **2017** | | | | | |
| ICML | Australia (Sydney, SYD) | Iceland (Reykjavík, KEF) | 54.6% | United States (Washington, JFK) | 44.6% |
| IJCNN | United States (Anchorage, Alaska, ANC) | Norway (Oslo, OSL) | 8.8% | United States (Washington, JFK) | -4.1% |
| IEEE CEC | Spain (San Sebastián, EAS) | Germany (Berlin, BER) | 6.0% | China (Beijing, PEK) | -19.3% |
| IEEE ICRA | Singapore (Singapore, SIN) | Norway (Oslo, OSL) | 47.4% | United States (Washington, JFK) | 31.9% |
| **2018** | | | | | |
| ECCV | Germany (Munich, MUC) | Russia (Moscow, SVO) | 6.8% | United States (Washington, JFK) | -27.5% |
| ICML | Sweden (Stockholm, ARN) | Russia (Moscow, SVO) | 0.7% | United States (Washington, JFK) | -38.9% |
| IJCNN | Brazil (Rio de Janeiro, GIG) | United Kingdom (London, LHR) | 32.8% | Brazil (Brasilia, BSB) | 2.3% |
| IEEE CEC | Brazil (Rio de Janeiro, GIG) | Ireland (Dublin, DUB) | 21.2% | Brazil (Brasilia, BSB) | 1.8% |
| IEEE CVPR | United States (Salt Lake City, Utah, SLC) | Iceland (Reykjavik KEF) | 19.2% | United States (Washington, JFK) | 1.8% |
| IEEE ICRA | Australia (Brisbane, BNE) | Norway (Oslo, OSL) | 60.9% | United States (Washington, JFK) | 45.5% |
| **2019** | | | | | |
| ICML | United States (Long Beach, California, LAX) | Finland (Helsinki, HEL) | 35.9% | United States (Washington, JFK) | 8.7% |
| IJCNN | Hungary (Budapest, BUD) | China (Beijing, PEK) | 15.0% | China (Beijing, PEK) | 15.0% |
| IEEE CEC | New Zealand (Wellington, WLG) | China (Beijing, PEK) | 55.4% | China (Beijing, PEK) | 55.4% |
| IEEE CVPR | United States (Long Beach, California, LAX) | Russia (Moscow, SVO) | 30.1% | China (Beijing, PEK) | 32.8% |
| IEEE ICRA | Canada (Montreal, Quebec, YUL) | China (Beijing, PEK) | 63.9% | China (Beijing, PEK) | 63.9% |

**Table 3.** Actual conference locations (country, town, airport) and optimal locations based on all countries (BOC) and based on paper submissions (BPS). Savings (%) in CO2 emissions for conference air-travelling to optimal locations, 2020 to 2021 (during COVID-19).

| Conf. Name | Actual location | Optimal location BOC | Savings in CO2 - location BOC | Optimal location BPS | Savings in CO2 - location BPS |
|---|---|---|---|---|---|
| **2020** | | | | | |
| ECCV | United Kingdom (Glasgow, GLA) - Virtual | Russia (Moscow, SVO) | 8.2% | China (Beijing, PEK) | 12.1% |
| ICML | Austria (Vienna, VIE) - Virtual | China (Beijing, PEK) | 5.3% | China (Beijing, PEK) | 5.3% |
| IJCNN | United Kingdom (Glasgow, GLA) | Estonia (Tallinn, TLL) | 5.5% | China (Beijing, PEK) | 0.0% |
| IEEE CEC | United Kingdom (Glasgow, GLA) | Estonia (Tallinn, TLL) | 4.0% | China (Beijing, PEK) | -4.0% |
| IEEE CVPR | United States (Seattle, SEA) - Virtual | Russia (Moscow, SVO) | 23.1% | China (Beijing, PEK) | 27.1B% |
| IEEE ICRA | France (Paris, CDG) - Virtual | Norway (Oslo, OSL) | 4.2% | United States (Washington, JFK) | -28.1% |
| **2021** | | | | | |
| ICML | Austria (Vienna, VIE) - Virtual | China (Beijing, PEK) | 29.4% | China (Beijing, PEK) | 29.4% |
| IJCNN | China (Shenzhen, SZX) - Virtual | China (Beijing, PEK) | 12.9% | China (Beijing, PEK) | 12.9% |
| IEEE CEC | Poland (Kraków, KRK) - Virtual | Estonia (Tallinn, TLL) | 1.6% | China (Beijing, PEK) | -5.9% |
| IEEE CVPR | United States (Nashville, Tennessee, BNA) - Virtual | Mongolia (Ulaanbaatar, UBN) | 36.9% | China (Beijing, PEK) | 37.6% |
| IEEE ICRA | China (Xi'an, XIY) - Hybrid | Iceland (Reykjavik, KEF) | 18.6% | United States (Washington, JFK) | 0.9% |

## 4  Discussion

The results of this research pose rethinking to conferences' organizers and participants to consider the true objectives of scientific meetings and to evaluate the trade-offs posed within this article. It is clear that green-conference alternatives need to be seriously considered. This work clearly indicated that by simply changing the conference location to an optimal one, an impactful decrease in the carbon emissions due to air-travelling can be accomplished.

A finite set of well-known Artificial Intelligence conferences was selected to point out this argument by drawing some initial conclusions. More accurate results would require the exact number of participants of each conference and their town of origin/trip-start at that time, which was not feasible to be retrieved. It is obvious that even if the optimal location was set as the conference organizing town, still some people would have to travel in order to attend.

Results indicate that virtual conferencing comes first in the list of green conferences' conduction ways, leading to 100% of CO2 savings, although people are skeptical, mainly due to poor virtual networking. Even though results are predictable, it is the first time that they are quantified. Quantifying our results, the conduction of the six Artificial Intelligence conferences in their actual location resulted in average CO2 emission tonnes equal to 1938.70 for ECCV, 5000.62 for ICML, 2010.89 for IJCNN, 1385.45 for IEEE CEC, 5636.50 for IEEE CVPR and 2837.68 for IEEE ICRA. Results rank IEEE CVPR at the top regarding air-travelling participation emissions. Considering the 2016 per capita CO2 emissions of the countries [31], the average cumulative CO2 emissions of the six conferences (18809.84 tonnes) correspond to the annual emissions of 1212, 2549, 2944, 3667, 4143, 8360, 9848 citizens in USA, China, Greece, France, Sweden, Brazil and India, respectively. Tables 2 and 3 indicate that while most submissions on IEEE CVPR were always from China, the conference was always held in the United States. This is an indication that conference locations are not eco-friendly chosen, definitely.

Based on the results of our research, by optimally deciding the conference's location, a decrease of up to 63.9% in gas emissions was reported. Therefore, an eco-friendly alternative is emerging, reporting disadvantages as well; as it can be observed from Tables 2 and 3, most submitted papers were always either from the United States or China, proportionally due to their population or due to being more scientifically active in the field of Artificial Intelligence. The latter will cause US (Washington) and China (Beijing) resulting always as the optimal locations and eventually be the only organizers of scientific events.

A limitation posed from the proposed research design is that not all data from all selected conferences through the predefined decade could be retrieved due to not well-organized proceedings. The focus of this work is to reveal a trend, therefore absolute values are of no particular purpose. However, future work is expected to retrieve the missing data. Moreover, one would assume that it is not feasible to implement the selection of location based on the paper submissions; this would mean that the conference location would be announced after the submissions, which contradicts what is happening so far, that is, based on the conference location, the submissions are mainly motivated. Overall, the aim of this work is to quantify the CO2 emissions for travel to specific AI conferences, analyze the factors that influence this travel and explore possible alternatives and ecological approaches. These approaches are suggested for discussion, although additional concerns should be addressed to practically implement them.

In order to conclude to a place to organize a conference, the committee's criteria may be the place of origin of its members, or touristic destinations that will attract people, etc. The aim of this work is to try for the first time to quantify air-travelling conferencing emissions and look at the footprint of AI conferences, excluding such factors. Future work could also include additional criteria based on the strategies followed by conference committees and could discuss feasible alternatives of green AI conferencing.

Regarding conference air-travelling emissions, on one hand, the International Panel of Climate Change (PCC) and the UNFCCC guidelines underline that CO2 emissions from international aviation are not accounted for in national totals and are reported separately [31]. Therefore, CO2 travelling emissions are not included in the limitation and reduction commitments of Annex I Parties under the Convention and the Kyoto Protocol and therefore no serious efforts are made to reduce them. On the other hand, the organizing towns benefit from the conference-related tourism to boost their local economy. One will wonder if this is politically and ethically correct. Moreover, airport-specific green rating frameworks need to be considered [32].

Artificial Intelligence community should be sensitized and take action, since within the scope of its research is among others to accelerate global efforts to protect the environment and conserve resources, by developing artificial intelligence prediction models. Redefinitions and reconsiderations on current conferences' conduction may be in need.

Researchers in the field of Artificial Intelligence should be aware of the impact of conference air-travelling to the environment and in the future to be able to decide on their carbon footprint individually. Moreover, organizing countries should propose fluctuating conference registration fees according to the participants' affiliation location so as to raise awareness.

Currently, many conferences are still held online. The latter is due to COVID-19 pre-existing conditions, which created infrastructure and experience for the organization of remote conferences, to facilitate travel and enhance green conferences. Due to COVID-19, conferencing mobility was frozen, becoming the catalyst of online communication, paving the way to a massive migration to the metaverse [33]. The concept of the metaverse lies in living in a higher-dimensional world, which goes beyond the physical world we live in. Building conference metaverses may be the trend of the future, so that participants could interact with past events, keep track of the entire conference sessions, being in the virtual space without the need to air-travel with all its consequences, economical, and ecological. It is believed that we have entered a new stage and very likely, there is no turning back. To this end, since the impact on networking is strong, we see that conferences are gradually returning to on-site participation offering at the same time the possibility for on-line participation, balancing between physical and metaverse conferencing.

## 5    Conclusions

In this work, for the first time it is quantified the impact of scientific conferences' air-travelling. Alternative ways for greener conferences towards reducing the global carbon footprint are examined. The focus is on the most popular AI conferences based on their scientific impact factor, their scale, and the well-organized proceedings

Our findings highlight that the virtual way is the first on the list of green conferences' conduction although there are serious concerns about it. Alternatives to optimal conferences' location selection have demonstrated savings on air-travelling CO2 emissions of up to 63.9%. After this research on the field of Artificial Intelligence conferences, we believe that similar research should be conducted for more scientific fields so as to create eco-friendly alternatives towards a global framework of a greener way for conferences conduction.

## Acknowledgement

This work was supported by the MPhil program "Advanced Technologies in Informatics and Computers", hosted by the Department of Computer Science, International Hellenic University, Kavala, Greece.